\documentclass[ twocolumn,aps,pre]{revtex4}
\usepackage{graphicx,color}
\begin{document}
   \title{\bf {Coupling  driven exclusion and diffusion processes on 
parallel lanes: boundary induced phase transitions and boundary layers}}
\author{Bappa Saha and  Sutapa Mukherji}
\affiliation{Department of Physics, Indian Institute of Technology,
Kanpur-208 016}
\date{\today}
\begin{abstract}
We study a driven many particle system comprising
 of two identical lanes of finite lengths. On one lane, particles hop 
diffusively with a bias in a specific direction. 
On the other lane, particles  hop in a specific 
direction obeying mutual exclusion rule. 
In addition, the two lanes are 
connected  with each other through exchange of particles 
with certain rules.  The system, at its two ends, is in contact 
with particle reservoirs which maintain specific
 particle densities at the two ends. 
In this paper, we study boundary-induced phase transitions 
exhibited by this system and predict  the phase diagram
 using the technique of  fixed point based  
boundary layer analysis. 
An interesting manifestation of the 
interplay of  two density variables associated with two lanes 
is found in the  shock phase in which the particle 
 density profile across the lane with unidirectional hopping 
shows a jump discontinuity (shock) from a low to 
a high density region. The density profile on 
the diffusion-lane never exhibits a shock. However, 
 the shock in the other lane gives rise to a discontinuity  
in the slope of the diffusion-lane density profile. 
We show how an approximate solution for the slope can be 
obtained in the  boundary layer analysis framework.  
\end{abstract}
\maketitle
\section{Introduction}
Asymmetric simple exclusion processes (ASEP) 
are simple non-equilibrium systems of driven 
interacting particles. These systems have been 
studied extensively in the past since  they show interesting 
non-equilibrium behavior such as boundary induced phase 
transitions \cite{krug}, 
spontaneous symmetry breaking\cite{evans1}, 
phase separation \cite{janowsky} etc. Further, 
ASEPs serve as simple models to understand 
intracellular transport processes
\cite{hirokawa,chou, howard}, 
traffic flow \cite{mukamel,helbing} etc. 
The simplest and the most extensively
studied ASEP involves 
  a one-dimensional lattice on  which particles  hop 
   in a specific direction. Particles obey the 
mutual exclusion rule that forbids two 
particles from occupying the same site. 
In the case of a finite lattice, the boundaries of the 
system are attached to reservoirs which maintain 
specific densities at the boundaries. 
Unlike equilibrium systems, these  
one-dimensional systems with short-range interaction 
 exhibit boundary induced bulk phase transitions 
in which  the boundary rates determine 
the bulk properties of the system. Distinct transport properties and 
diverse shapes of particle distribution profiles at different bulk phases 
result from the subtle interplay of the particle current, 
the inter-particle interaction and the boundaries of the system. 

In intracellular transport processes, various kinds of motor 
proteins hop on the biopolymers to transport different 
cellular constituents to definite locations inside  the cell. 
In order to have a closer resemblance with these 
biological transport processes, different kinds of ASEPs 
have been proposed up to now.  
The specific ASEP  that we consider here 
consists of two one-dimensional finite 
lanes (lattices). On one lane, particles have diffusive motion 
with a bias. We name this 
lane as the diffusion-lane.  
On the other lane, to be called as  the ASEP-lane, 
 particles hop in a specific direction 
obeying the mutual exclusion rule.
  In addition, between the lanes,  there are 
  particle exchange processes  
which lead to a \textquoteleft transverse flux\textquoteright\  
of particles. 
The indirect mutual interaction between 
the lanes caused by particle exchange  leads 
to a nontrivial coupling between the particle-densities on 
the two lanes and 
as a consequence of this, the phase diagram associated with the 
ASEP-lane  is expected to be significantly different from that 
 of a single lane ASEP. The two-lane model we consider here 
 was proposed earlier in  order to model   intracellular 
transport processes realistically.
The ASEP lane mimics the hopping of 
the  motor particles along 
the biopolymers. The  diffusive lane in this model 
replaces the environment in which the motor particles 
diffuse during time intervals when they are not 
attached to  the biopolymers on which they hop 
\cite{klumpp}. Later, a similar model was proposed in the 
context of extraction of membrane tubes by motor particles 
\cite{tailleur}.

In the past, the  phase diagram for this model has been obtained 
under zero \textquoteleft transverse flux\textquoteright\  
 condition  between the lanes
\cite{evans2}.  The boundary conditions maintained by the 
boundary reservoirs are also 
assumed to respect  this zero \textquoteleft 
transverse flux\textquoteright\   condition.
 This  condition ensures 
that despite the  exchange processes, the average number of particles on 
a lane is conserved and as a result, the average particle density  
 across the bulk of the 
lane remains constant (a flat profile). 
Further, this zero 
\textquoteleft transverse flux\textquoteright\  condition 
leads to a functional relation 
between the two density variables and thus 
the problem  reduces to a single lane problem 
with an effective current density  that is 
different from the hopping current of the 
usual single lane ASEP \cite{derrida0}. 
The nature of the bulk  profile 
and  the  locations of the boundary layers  
are also predicted through a fixed point based 
boundary layer analysis \cite{vandana}.
 In this article, allowing a nonzero 
\textquoteleft transverse flux \textquoteright\  
 between the lanes and unconstrained boundary densities,
 we attempt to obtain general solutions for the steady-state 
profiles  for this coupled two lane problem.  

Due to their simplistic nature, much progress has been made 
on the simplest single-lane ASEP with open boundaries.  
  In addition to exact solutions \cite{derrida0,derrida,schuetz}, 
there exist other studies 
based on domain wall dynamics 
\cite{straley,santen}, maximum current principle 
\cite{krug}, mean-field 
theory \cite{derrida0} etc which provide significant 
insights on boundary induced phase transitions in this system.  
Some of these approaches have been generalized  
to systems with 
different inter-particle interactions \cite{hager},
 disordered hopping rates \cite{stinchcombe}, 
 particle adsorption-desorption processes
\cite{parameggiani,evans3}  etc.  Apart from these, 
techniques  based on boundary-layer analysis have 
also been developed to study   phase 
transitions  in these systems  \cite{smsmb,smint}. 
 This analysis shows 
that the origin of  phase transitions and the  nature  
of the phase boundaries can be understood 
from the shapes of the   steady-state  density 
profile and the location of its  boundary layer 
regions  under different boundary 
conditions. 
Processes with multi-species \cite{popkov2species,smpre}, 
multi-lanes \cite{popkovsalerno,melbinger,
schiffmann,evans2,vandana}  are more 
complex since, in general, these 
processes involve two or more density variables. 
In order to study these complex processes, 
  boundary layer analysis of reference \cite{smsmb}, has been 
  further generalized in \cite{smpre}
   by developing a fixed point based boundary 
  layer analysis. This work also provides a holographic 
  interpretation of the method which allows predictions
about the bulk profile and bulk phase transitions
 through a  fixed point analysis 
  of the boundary layer region.
 These studies  motivate us to obtain a 
systematic way to find out how, 
for the present two-lane problem, the boundary 
layer or the bulk profile
 changes with the boundary conditions.  
Using boundary layer analysis,  we 
 solve approximately the steady-state differential equation 
describing the density profile.

 From this study, we conclude that 
while the average density profile of the diffusion-lane remains 
less  sensitive to the boundary densities, 
the density profile on the ASEP-lane undergoes 
significant changes as the boundary densities 
 are changed.  The present 
 problem has similarities with single lane ASEP 
with additional particle adsorption-desorption processes 
(Langmuir Kinetics).
 However, in our case, the particle 
exchange happens with a lane which is also dynamically evolving.  
This is reflected in the  phase diagram whose 
overall structure  is different from 
that of the single lane ASEP, although 
 similar to the single lane ASEP, here also 
we find three major phases such as low-density, 
high-density and shock phases. We follow the conventional rule for naming 
these phases i.e.  in the low- and high-density 
phases, the major part of the density profile has a value 
less than or greater than $0.5$, respectively, and  in 
the shock phase,  there is a localized shock 
(discontinuity) in the 
bulk part of the profile separating a low-
and a high-density region.
We show that the diffusion-lane cannot have a shock. 
However, the  shock in the ASEP-lane  affects the 
 diffusion-lane's profile 
by creating a discontinuity in the slope of its profile.          
The boundary-layer method provides an approximate 
 but a systematic way to study the slope analytically.

The paper is divided into   following sections. 
In section II, we introduce the model and  obtain the 
corresponding mean-field equations.  In section III, 
we discuss the boundary-layer analysis for the model 
in detail. Results and the phase diagram are discussed in 
Section IV.
In section V, we present a summary of our work.

\section{Definition of the model and mean-field description}
\subsection{The model }
 The model comprises of two lanes each having  length $L$ and   
 $N$ number of sites  with lattice spacing $a$ $(L=Na)$. 
On one of the lanes, referred  as the diffusion-lane,
particles move with biased diffusion with rates $D^+$ and  
$D^-$ ($D^+\neq D^-$) towards their right and left, respectively.  
On the other  lane, referred as the  ASEP-lane, particles  
move towards the neighboring site on the 
 right  respecting  exclusion principle. We assume that a 
particle hops to the target site with unit 
probability if the target site is empty. 
In addition, there are particle exchange processes from
 one lane to the other. 
A particle from a diffusion-lane can leave its 
lane and occupy a site of the ASEP-lane provided 
the site is empty. We assume that this process 
takes place at rate $\omega_{\rm a}$. Similarly, a particle from 
the ASEP-lane can detach 
itself from its lane and occupy a site  on the diffusion-lane 
at a rate $\omega_{\rm d}$. 

Finally, it is assumed that both the lanes  are connected at the 
boundaries to particle reservoirs which maintain specific 
densities at the two ends of the lanes.

\subsection{Mean-field equations}
We denote the particle-occupancies of the $i$-th site of   the 
ASEP- and diffusion-lane 
by  $\tau_i$ and  $\sigma_i$, respectively. $\tau_i=1,\ 0
(\sigma_i=1,\ 0)$  imply that the 
 $i$th site of the ASEP-lane (diffusion-lane) is occupied  or unoccupied, 
respectively. In terms of these variables, 
the discrete Master equations 
describing the particle dynamics have the following 
forms \cite{evans2}
\begin{eqnarray}
&&\frac{{\rm d}{\tau}_i}{{\rm d}t} =-\tau_i(1-\tau_{i+1})+ 
\tau_{i-1}(1-\tau_i)-\nonumber\\
&&\tau_i\omega_{{\rm d}} + \omega_{{\rm a}}\sigma_i(1-\tau_i)   \label{1}\\ 
&&\frac{{\rm d}{\sigma}_i}{{\rm d}t} = D^-\sigma_{i+1}-D^+\sigma_i
+D^+\sigma_{i-1}-\nonumber\\
&&D^-\sigma_i +\tau_i\omega_{{\rm d}}-\omega_{{\rm a}}\sigma_i(1-\tau_i). \label{2}
\end{eqnarray}
A  statistical 
averaging of these master equations along  with the assumption that 
the occupancy variables are 
uncorrelated i.e. $\langle \tau_i \tau_{i+1}\rangle=
\langle \tau_i\rangle \langle \tau_{i+1}\rangle$ leads to 
the mean-field equations. Further, we go over to the 
continuum limit,  which implies 
$N\rightarrow \infty$, $a\rightarrow 0$ 
with  $L=Na$ finite. For simplicity, we choose the 
lattice size to be unity ($L=1$). 
In the continuum limit,  mean-field 
equations under  steady-state condition  
 ($\frac{\partial <\tau_i>}{\partial t'}=
\frac{\partial <\sigma_i>}{\partial t'}$=0, where 
$t'$ is the appropriate rescaled time) are  
\begin{eqnarray}
\frac{\epsilon}{2} \frac{\partial ^2\tau}{\partial x^2}-
\frac{\partial \tau}{\partial x}(1-2\tau)-\Omega_{{\rm d}}\tau + 
\Omega_{{\rm a}}\sigma(1-\tau)
 &=& 0 \label{3}\\
\epsilon D_{\sigma}\frac{\partial ^2\sigma}{\partial x^2}-
v\frac{\partial \sigma}{\partial x}+\Omega_{{\rm d}}\tau-
\Omega_{{\rm a}}\sigma(1-\tau) 
&=& 0, \label{4}
\end{eqnarray}
Here $\Omega_{{\rm a}}=\omega_{{\rm a}}N$,  $\Omega_{{\rm d}}=\omega_{{\rm d}}N$,  
$\epsilon = 1/N$, $v=D^+-D^-$,
$D_{\sigma}=(D^++D^-)/2$ and $\tau(x)$ and $\sigma(x)$ are 
the average densities at position 
 $x$ on the ASEP- and diffusion-lane, respectively. 
In order to obtain the continuum version from the discrete one, 
 a Taylor-expansion  as $\langle\tau_{i\pm 1}\rangle=
\tau(x)\pm \epsilon\frac{{\rm d}\tau(x)}{{\rm d}x}+
\frac{\epsilon^2}{2} \frac{{\rm d}^2\tau}{{\rm d}x^2}$ and the 
same for $\langle\sigma_{i\pm 1}\rangle$ 
have been  done. The second order derivatives in (\ref{3}) and 
(\ref{4}) originate from the second order terms in the Taylor expansion. 
These second order derivative terms play an 
important role in the 
boundary-layer analysis discussed 
below. In order to obtain the steady-state 
density profiles, one requires to solve 
the differential equations with given boundary densities 
maintained by the boundary reservoirs. 
We assume that the boundary densities at left end and right
end are $\alpha$ and $\gamma$ i.e. 
 $\tau(x=0)=\sigma(x=0)=\alpha$ and 
$\tau(x=1)=\sigma(x=1)=\gamma$.

The total horizontal current in the system is 
\begin{equation}
J_{{\rm tot}} = J_{\tau}+J_{\sigma} = 
\tau(1-\tau)+v\sigma, \label{5}
\end{equation} where
$J_{\tau}=\tau(1-\tau)$, $J_{\sigma}=v\sigma$ are
  the particle currents in the ASEP- and 
diffusion-lane, respectively. 
In addition, there  is   also a 
\textquoteleft transverse particle flux\textquoteright, 
$T=\Omega_{{\rm a}}\sigma(1-\tau)-\Omega_{{\rm d}}\tau$,
from one lane to the other arising from the particle 
exchange processes.  

 An assumption that the net \textquoteleft transverse 
flux\textquoteright\ is zero ($T=0$) 
implies a  relation between the two densities 
$\sigma=\frac{\Omega_{{\rm d}}}{\Omega_{{\rm a}}} \frac{\tau}{1-\tau}$. 
Earlier studies \cite{evans2,vandana}, based on this  assumption, 
 were largely simplified because of this relation.     
In this work, we assume $T\neq 0$. In order to 
obtain the steady-state density profiles, 
one has to, therefore,  solve two coupled nonlinear 
equations for two densities 
under  various  boundary conditions.  In the following, 
we discuss the boundary-layer analysis  of this
general system.

\section{Boundary-layer analysis}
In  boundary-layer analysis, 
based on the relevance of various terms 
at different length scales, 
one usually  breaks a 
differential equation into different parts \cite{jd_cole}. 
The  approximate differential equations obtained thereby, 
 are solved and the  solutions valid over different length scales 
are matched asymptotically to obtain a uniform approximate 
solution of the complete original 
differential equation over the entire domain 
of interest. 
The strategy of focusing on the relevant parts of the 
differential equation at different length scales 
often simplifies the problem even if the starting equation has 
a complicated structure.

\subsection{Outer or bulk equation}
In the thermodynamic limit, $N\rightarrow \infty$ 
($\epsilon\rightarrow 0$),  it appears natural to ignore 
the second order derivative terms of equations (\ref{3}) and (\ref{4}).   
The resulting differential equations, known commonly as outer equations, 
are 
\begin{eqnarray}
 \frac{{\rm d}\tau_{\rm out}}{{\rm d}x}(1-2\tau_{\rm out})+
\left(\Omega_{{\rm d}}\tau_{\rm out}-\Omega_{{\rm a}}
\sigma_{\rm out}(1-\tau_{\rm out})\right) = 0 \label{out1}\\
 v\frac{{\rm d}\sigma_{\rm out}}{{\rm d}x}-\left(\Omega_{{\rm d}}\tau_{\rm out}-
\Omega_{{\rm a}}\sigma_{\rm out}(1-\tau_{\rm out})\right)=0, \label{out2}
 \end{eqnarray}
 where, a subscript ``${\rm out}$'' is introduced to identify the 
densities as solutions of the outer equations. 
The solutions of these equations, referred in the following 
 as the outer solutions or bulk solutions, 
describe the major part of the density profiles. 
 It is straightforward to see that  the bulk solutions  
satisfy the condition $\frac{{\rm d}J_{\rm tot}}{{\rm d}x}=0$  
which means $J_{\rm tot}$ is  constant over
the region where bulk solutions for both $\tau$ and $\sigma$ together 
 prevail. It must, however,  be noticed that a 
solution of a  first order equation as mentioned above can  satisfy 
only one   boundary condition. 

\subsection{Inner or boundary-layer equation}
\begin{figure}[htbp]
  \centering
   \includegraphics[height=.35\textwidth]{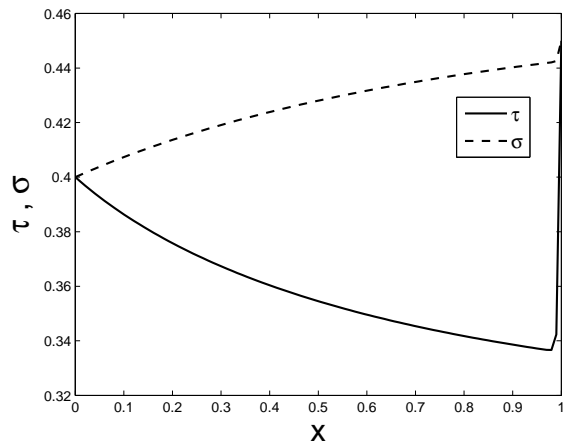}  
\caption{Density profiles for  $\alpha= 0.4$ and $\gamma= 0.45$. 
Values of other parameters are $\Omega_{{\rm d}}=\Omega_{{\rm a}}=0.2$, 
$\epsilon=0.002$, $D_\sigma=0.6$ and $v=0.4$.}
\label{fig:samplefig}
\end{figure}

Since the density profile 
  must  also satisfy the other boundary 
condition, the entire density profile cannot  
be described  by the outer solution  alone.
 This implies that the density profile  must have  
another distinctly different 
part, to be called as inner solution or boundary-layer solution,
satisfying either of the following conditions.
(a) In case the outer solution 
satisfies one boundary condition, 
 the inner or the boundary layer 
 solution  appears near the other boundary  
in order to satisfy the  boundary condition there. 
For example, in figure (\ref{fig:samplefig}), 
the inner solutions for both $\tau$ and $\sigma$ 
  satisfy the boundary condition at $x=1$
while   their  outer solutions satisfy  the boundary 
condition at $x=0$. 

\begin{figure}[htbp]
  \begin{center}
  (a) \includegraphics[width=.4\textwidth,clip,
angle=0]{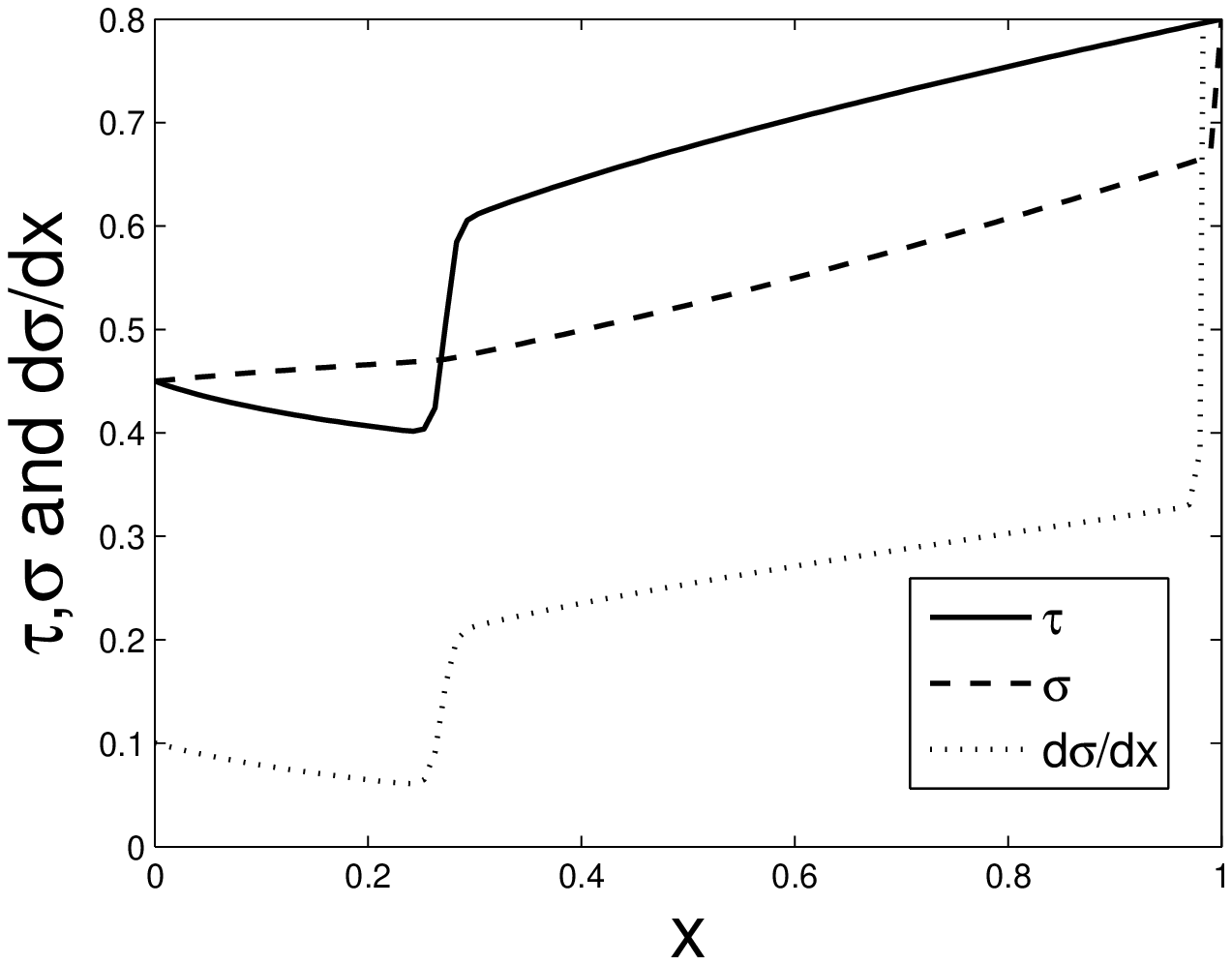}\\
(b)   \includegraphics[width=.4\textwidth,clip,
angle=0]{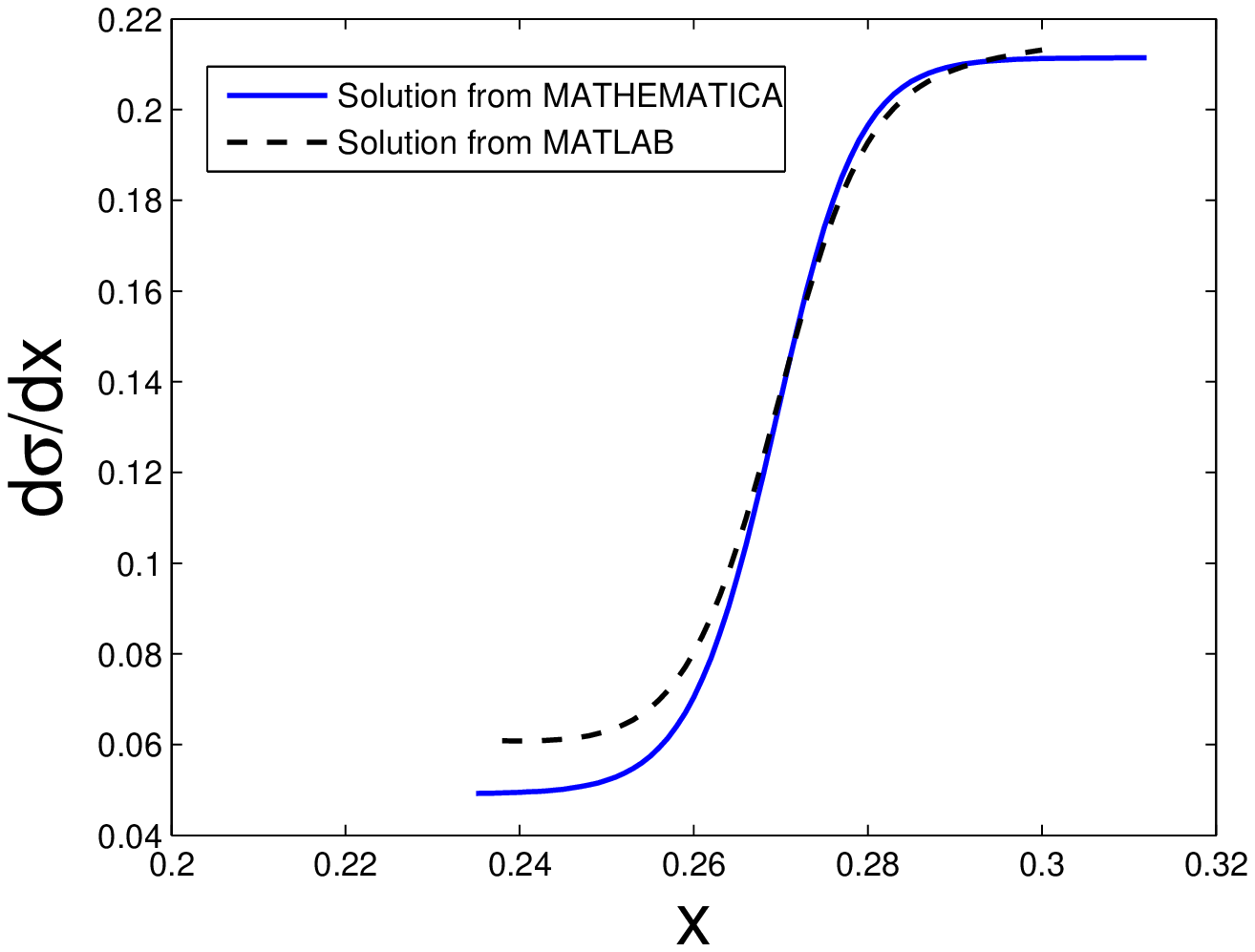}
\caption{(a) Density profiles and the 
slope of $\sigma$-profile  for $\alpha= 0.45$ and 
$\gamma= 0.8$.   (b)  Slope of the 
 $\sigma$-profile from analytical and numerical solutions.
Other parameter values are same as those in figure (\ref{fig:samplefig}).}
\label{fig:shock}
  \end{center}
\end{figure}

(b) The  inner solution  may be located 
somewhere in the interior  of the system with two 
outer solutions appearing on  two sides of this 
 solution (shown by the solid line in figure (\ref{fig:shock})). Here 
the inner solution
acts like a domain wall  separating two outer 
solutions with high and low density values.
Two disjoint outer solution parts satisfy 
the two boundary conditions. 
Thus although this solution is referred 
as inner or boundary-layer solution, it   may not 
necessarily appear near the boundary.  In a similar manner, 
the outer or the bulk solution  may 
start from the boundaries and need not always 
 be confined to the interior  of the system.

The corresponding differential equation (known as the 
inner equation)
 can be found out upon  expressing (\ref{3}) and (\ref{4}) 
in terms of  a rescaled variable  
$\tilde x=\frac{x-x_c}{\epsilon}$, where $x_c$ represents the 
location of the boundary layer,  and then 
 ignoring terms of $O(\epsilon)$. 
The inner or the boundary layer solution 
 is a rapidly varying solution  
since it  appears over a length scale $\epsilon$ and 
 becomes sharper as $\epsilon\rightarrow 0$. The outer or the 
bulk solution on the other hand is a much slowly 
varying solution.

Since the    two independent 
bulk solutions appearing on  either side 
of the inner solution in figure (\ref{fig:shock}) 
are the solutions of a first order outer equation, each 
outer solution has  
one integration constant. These constants are fixed by 
claiming that each bulk solution satisfies one boundary condition. 
The boundary layer located at $x=x_c$, on the other hand, 
merges to the outer solutions smoothly 
as $\tilde x\rightarrow \pm \infty$.
The constraints on the inner solution are different 
when the inner solution appears near one boundary. 
In this case, the inner solution at one end satisfies
 the boundary condition and on the other end merges to 
the bulk (see figure (\ref{fig:samplefig})). 
More explicitly,  
an inner solution  near $x=x_c\approx 0$ or  $x=x_c\approx 1$, 
approaches the bulk as 
$\tilde x\rightarrow \infty\ {\rm or} -\infty$, respectively.   
In any case, since the boundary layer has to satisfy 
more than one condition, it is expected that the 
inner equation is a higher order equation in comparison 
with the outer equation. 
In terms of $\tilde x$ introduced before,   
we have second order inner equations as shown  below. 
\begin{eqnarray}
  D_{\sigma}\frac{\partial ^2\sigma_{\rm in}}{\partial \tilde{x}^2} 
= v\frac{\partial \sigma_{\rm in}}{\partial \tilde{x}},\label{in1} 
\end{eqnarray} 
\begin{eqnarray}
  \frac{1}{2}\frac{\partial ^2\tau_{\rm in}}{\partial \tilde{x}^2}
 = \frac{\partial \{\tau_{\rm in}(1-\tau_{\rm in})\}}
{\partial \tilde{x}}. \label{in2}
\end{eqnarray}

In the following analysis, the density profile will be primarily 
characterized by the location and the slope  of the 
boundary layer  as well as  the bulk solutions. These 
properties depend on whether  a particular 
boundary-layer solution is able to satisfy all the 
constraints under  given boundary conditions.  
 
\subsection{The boundary layer, the bulk solutions and  matching of the two}
Bulk solutions can be obtained by solving the nonlinear coupled 
equations (\ref{out1}) and (\ref{out2}). It appears
 that it is more convenient to solve the equations 
numerically with given boundary conditions. 

As we shall show below, for our purpose, it is sufficient 
to have a knowledge  about the signs of the slopes of  the 
 densities  with $x$. The slope
 of the outer solutions can be specified 
conveniently in $\tau-\sigma$ plane 
(see figure {\ref{fig:bulkslope}). 
It is clear  from equations (\ref{out1}) 
and (\ref{out2}) that, the line  
$\sigma_{{\rm out}}=\frac{\Omega_{{\rm d}}}{\Omega_{{\rm a}}} \frac{\tau_{{\rm out}}}{(1-\tau_{{\rm out}})}$ plays
an important role in  deciding the slopes.
\begin{figure}[htbp]
 (a)  \includegraphics[width=.35\textwidth]{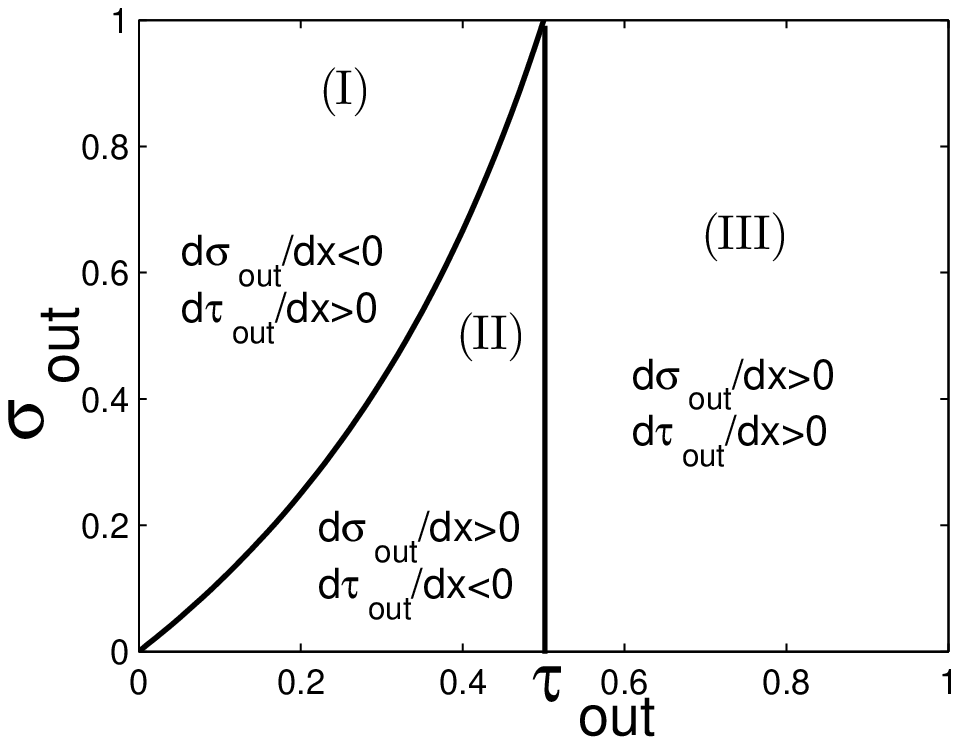}
(b)\includegraphics[width=.35\textwidth]{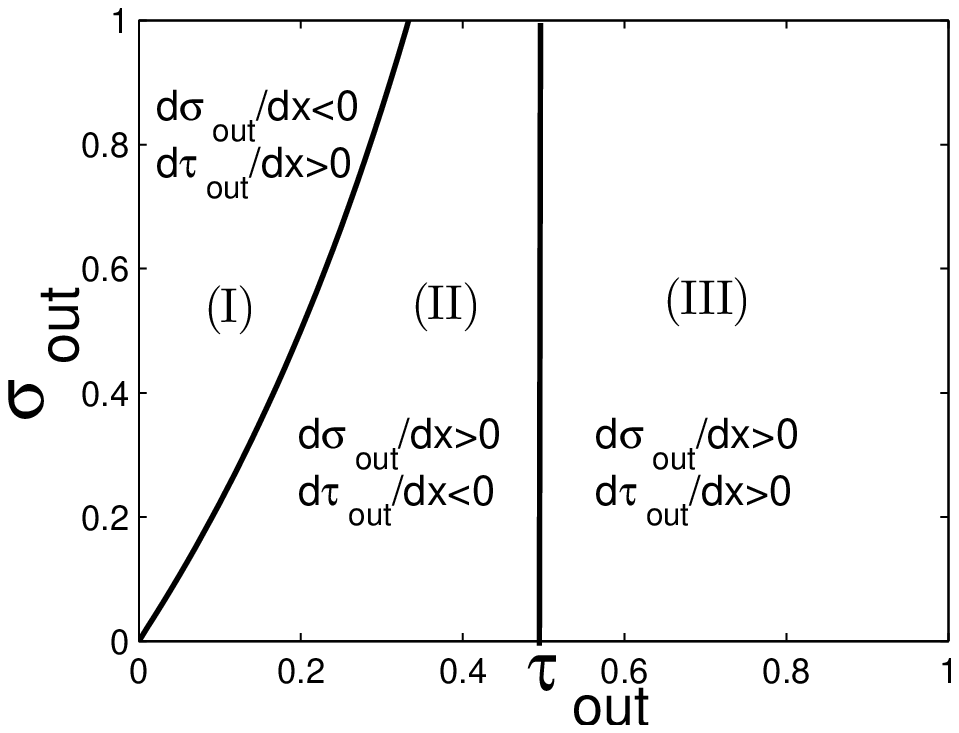}
\caption{The signs of the 
slopes of outer  solutions for different  values 
of $\tau$ and $\sigma$.
The solid curve corresponds to $\sigma_{\rm out}=
\frac{\Omega_{{\rm d}}}{\Omega_{{\rm a}}}\frac{\tau_{\rm out}}{1-\tau_{\rm out}}$. 
(a) for $\Omega_{{\rm d}}=\Omega_{{\rm a}}=0.2,\ v=0.4$ and 
(b) for $\Omega_{{\rm d}}=2\Omega_{{\rm a}}=0.2,\ v=0.4$.}
\label{fig:bulkslope}
\end{figure}

The inner equations, (\ref{in1}) and (\ref{in2}),  
are decoupled and can be solved exactly.
To obtain the inner solution for the density 
on the diffusion-lane, we  integrate
 (\ref{in1}) once and find
\begin{eqnarray}  
\frac{{\rm d}\sigma_{{\rm in}}}{{\rm d}\tilde{x}} &=& \frac{v}{D_{\sigma}}\sigma_{{\rm in}}+ 
\frac{C_\sigma}{D_{\sigma}}, \label{in-int-1}
\end{eqnarray}
where $C_\sigma$ is the integration constant. Since the  
boundary layer is expected to saturate to the bulk value
($\frac{{\rm d}\sigma_{{\rm in}}}{{\rm d}\tilde{x}}=0$) in the appropriate limit,
it requires $C_\sigma=-v\sigma_0$, where $\sigma_0$ is the 
value of the bulk density to which the boundary layer saturates. 
In terms of $\sigma_0$, the solution of (\ref{in-int-1})
is 
\begin{eqnarray}
\sigma_{{\rm in}} = \frac{D_{\sigma}}{v}\left[ \exp (\frac{v}{D_{\sigma}}
 (\tilde{x}+k_0))\right]+\sigma_0, \label{in-sig-soln}
\end{eqnarray}
where $k_0$ is the second integration constant. 
If the boundary layer appears near $x=x_c\approx 0$, the saturation to the bulk density 
is expected in the $\tilde x\rightarrow \infty$ limit. 
The exponential solution, diverging as $\tilde x\rightarrow \infty$,
 however, can never saturate 
to the bulk. This restricts the boundary-layer solution for $\sigma$ to  
only  $x=1$ with  the slope of the boundary layer at $x=1$ being  
governed by the equation 
\begin{eqnarray}
\frac{{\rm d}\sigma_{{\rm in}}}{{\rm d}\tilde{x}}=\left(\gamma - 
\sigma_0 \right)\frac{v}{D_{\sigma}}.
\end{eqnarray}
Through this observation, our question about 
the  location of the boundary layers is partly solved. 
This feature, in addition, 
is  useful for us since we  now know that,
irrespective of the nature of  $\tau$-profile and the 
values of the boundary densities,  the outer solution of 
$\sigma$ must satisfy the boundary condition at $x=0$ 
and this outer solution must continue until the other 
boundary where the boundary condition is satisfied 
 by a narrow boundary layer as given in equation  
(\ref{in-sig-soln}). 

In a similar way, the inner solution for the 
 ASEP-lane can be obtained 
by integrating ({\ref{in2}) once. As done before for the 
diffusion-lane, here also 
the integration constant is fixed 
by demanding saturation of the inner solution to the bulk density 
$\tau_0$. The inner equation thus obtained is 
\begin{equation}
\frac{1}{2} \frac{{\rm d}\tau_{{\rm in}}}{{\rm d}\tilde{x}} = 
\tau_{{\rm in}}(1-\tau_{{\rm in}})+C_\tau = -(\tau_{{\rm in}}-\tau_0)(\tau_{\rm in}-1+\tau_0),
 \label{inn-tau-1}
\end{equation} 
where $C_\tau=-\tau_0(1-\tau_0)$ is the integration constant.

The  fixed 
points of this equation are 
$\tau_{\pm}^*=\frac{1\pm\sqrt{1+4C_\tau}}{2}=(1-\tau_0),\tau_0$
(We have chosen 
$\sqrt{1+4C_\tau}=1-2\tau_0$).
  A linear stability 
 analysis about these fixed points shows that $\tau_+^*$ 
is stable (unstable) if $\tau_0<0.5\ (\tau_0>0.5)$. 
Similarly, $\tau_-^*$ is stable (unstable) if $\tau_0>0.5\
 (\tau_0<0.5)$. 
\begin{figure}[htbp]
  \centering
   \includegraphics[height=.35\textwidth]{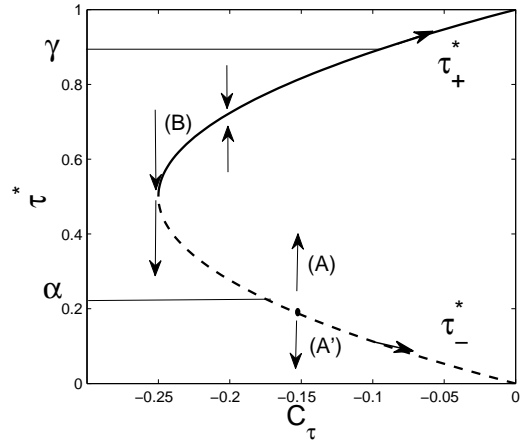}  
\caption{Fixed points of equation (\ref{inn-tau-1}) are 
plotted as functions of $C_\tau$ for $\tau_0<0.5$. 
Solid and dashed 
curves are referred in the text as upper 
and lower fixed point branches.
Vertical lines with arrows are the flow trajectories. 
Arrows on the upper and lower 
fixed point branches indicate that the  
outer solution is an increasing or decreasing function of 
$x$ for $\tau>0.5$ and $\tau<0.5$, respectively. 
Positive and negative slopes of the outer solution 
 correspond to region III and 
region II of figure (\ref{fig:bulkslope}), respectively.}
\label{fig:fixedptfig}
\end{figure}
Figure  (\ref{fig:fixedptfig}) shows the two fixed point branches along with 
 the flow behavior indicated through vertical upward or downward arrows.  

Equation  (\ref{in2}) can be also solved 
explicitly. The solutions are 
\begin{eqnarray}
\tau_{{\rm in}} &=& \frac{1}{2} +\frac{(1-2\tau_0)}{2}
\coth \left[ (\tilde{x}+k_1)(1-2\tau_0)\right]  \ {\rm or}   \nonumber \\
\tau_{{\rm in}}   &=& \frac{1}{2} + \frac{(1-2\tau_0)}{2} \tanh 
\left[(\tilde{x}+k_1)(1-2\tau_0)\right]. \label{inn-tau-sol}
\end{eqnarray}
 From the nature of these solutions 
in the $\tilde x\rightarrow \pm\infty$ limit, it is 
clear that  while the flow trajectories from 
the lower to the upper fixed point branch 
represent the $\tanh$ solutions,  the trajectories  
approaching the upper fixed point branch from above 
or moving further below  from the lower fixed point branch  
represent $\coth$ solutions. 
Although we have  explicit  analytical solutions
 for the boundary layers,
these are  not absolutely required for 
 predicting the qualitative shape of the 
density profile  and the location of  
the boundary layer under different boundary conditions.

\section{Phase diagram and phase boundaries}
Here we  primarily use the fixed point 
diagram (figure \ref{fig:fixedptfig}) and the 
diagram indicating the slopes 
 of the outer solutions  (figure \ref{fig:bulkslope}) 
to  predict  shapes of the density profiles under 
different boundary conditions. 
In order to do this, the following aspects  
may be useful. (i) In general, the 
boundary conditions $\tau(x=0)=\alpha$ and 
$\tau(x=1)=\gamma$ can be 
marked by two horizontal lines on the fixed point diagram.  
As shown in figure (\ref{fig:fixedptfig}), these lines 
intersect the fixed point branches at given points.
(ii)  The major part of the density profile is  
described by the outer solutions whose 
decreasing or increasing nature  with $x$ can be 
 predicted with the help of figure 
(\ref{fig:bulkslope}). The sign of 
the slope of the outer solution 
can also be conveniently indicated with 
arrows on the fixed point curves ( see figure \ref{fig:fixedptfig}).  
For example, arrows on upper and
 lower fixed point branches 
indicate that  the density 
in the outer solution increases or decreases with $x$, 
respectively. 
The decreasing nature of the outer solution 
for   $\tau<0.5$ corresponds to region II in 
figure (\ref{fig:bulkslope}). It can be argued that all the
$\tau<0.5$ cases that we have considered in the 
following correspond to region II of figure  (\ref{fig:bulkslope}).   
(iii) As mentioned earlier, the arrowed vertical lines represent 
different kinds of  boundary-layer solutions obtained  
earlier. The direction of the arrow in a vertical line 
of figure (\ref{fig:fixedptfig}) indicates where 
the boundary layer (represented by the vertical line)
 approaches as $\tilde x \rightarrow \infty$. 
As an example, the boundary layer in 
figure (\ref{fig:samplefig}) is of   
$\tanh$ type and it is 
represented by, say, a vertical line (A) 
in figure (\ref{fig:fixedptfig}). 
It must, however, be noted that 
the boundary layer of figure 
(\ref{fig:samplefig}) is just a part of 
the vertical line 
whose saturation to $\tau_0$ in the 
$\tilde x\rightarrow -\infty$ is apparent from 
 the density profile shown in the figure 
 but the saturation to $1-\tau_0$ 
as $\tilde x\rightarrow \infty$ happens at $x>1$ 
which is much beyond our physical region. 
In this case, the boundary layer satisfies 
the boundary condition before 
saturating to a higher value as $\tilde x\rightarrow \infty$.

While constructing the density profile for given boundary 
conditions, one may start   from, say, $x=0$ boundary 
and construct the density profile until $x=1$.   
Boundary conditions can be satisfied by  boundary layers or  
outer solutions. However, the entire construction of the density 
profile by combining the boundary layer and the outer solution 
parts must be consistent with  the flow trajectories and the sign of
 the slope of the outer solution. 

The phase diagram is broadly divided into two parts, $\alpha<0.5$ and 
$\alpha>0.5$. 
\subsection{$\alpha<0.5$}
\subsubsection{Low-density (LD) phase}

We begin with small values of  $\alpha$ and $\gamma$ 
such that both the horizontal lines 
corresponding to  $\tau=\alpha$ and $\gamma$
on the fixed point diagram 
intersect the lower fixed point branch. Since for the 
$\sigma$-profile,
 the boundary layer can  appear only near $x=1$, the 
boundary condition at $x=0$ is fulfilled by the outer solution 
which extends up to  the other boundary.  For the $\tau$-profile, 
directions of arrows on the vertical lines (see figure \ref{fig:fixedptfig})
indicate  that the boundary condition 
at $x=0$ cannot be satisfied by a boundary layer. This is because a
 boundary layer at $x=0$ must be represented by a vertical line  as (A), 
satisfying the boundary condition at $x=0$ and finally meeting the
 upper fixed point branch. Thus  near $x=0$, the   boundary layer 
 saturates to a  bulk solution which  always has a value  higher 
 than  $0.5$.  It follows from the fixed point 
 diagram that a boundary condition  $\tau(x=1)=\gamma<0.5$ can never 
 be satisfied by such a  bulk solution or by a boundary layer connecting this
  bulk solution.
  
Since the possibility of a boundary layer at $x=0$ is ruled out, 
 the density profile must have  an  outer solution that satisfies the 
boundary conditions   at $x=0$ and 
extends up to $x=1$ with a negative slope. 
This amounts to moving along the lower fixed point branch 
in the direction of the arrow starting from the boundary value $\alpha$. 
We stop at a given density which should be the value of the outer 
solution at $x=1$.  
The outer solution can then be followed by  a boundary layer
which finally satisfies the boundary condition at $x=1$.
The  nature of the boundary layer at $x=1$ is 
now decided in the following way.  If $\gamma>\alpha$, 
the boundary condition at $x=1$ can be satisfied only
through a $\tanh$ kind of a boundary layer indicated by 
the vertical line (A) in figure (\ref{fig:fixedptfig}). If 
$\alpha>\gamma$, there can be boundary 
layers of $\tanh$ or $\coth$ type depending on the value of 
$\tau_{\rm out}(x=1)$. For 
$\tau_{\rm out}(x=1)>\gamma$, a $\coth$ type boundary
 layer of negative slope appears at $x=1$ 
 (vertical line (A')). 
A $\tanh$ type boundary layer appears at $x=1$ if 
$\tau_{\rm out}(x=1)<\gamma$
(vertical line similar to (A)). 
While this $\tanh$ type boundary layer saturates 
to $\tau_{\rm out}(x=1)$} as $\tilde x\rightarrow -\infty$, 
the saturation 
 to $1-\tau_{\rm out}(x=1)$ as $\tilde x\rightarrow \infty$ 
happens much beyond the physical boundary at $x=1$.   
The dashed line between phases IA and IB in figure (\ref{fig:phases1}) 
indicates the change in the slope of the boundary layer at $x=1$.
This  phase boundary is, therefore,   the solution of the equation  
$\tau_{\rm out}(x=1)=\gamma$
where $\tau_{\rm out}(x=1)$ is a function of $\alpha$.  
We call this phase as the low-density phase since 
the value of the bulk density  remains 
within the lower half of the fixed point diagram.  
\begin{figure}[htbp]
\centering
   \includegraphics[width=.45\textwidth]{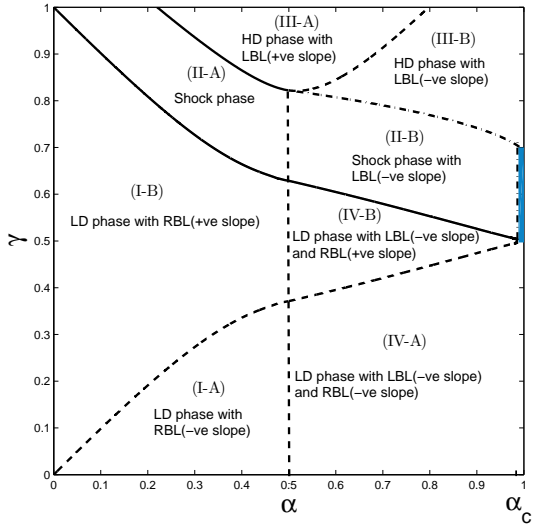}
\caption{Phase diagram for $\epsilon = 0.002$, 
$D_{\sigma} = 0.6$, $v=0.4$, $\Omega_{{\rm d}}=\Omega_{{\rm a}}=0.2$. 
As the dot-dashed lines are approached from the shock 
phase, the shock height vanishes continuously. RBL and LBL stand for 
right boundary layer (boundary layer at $x=1$) and left 
boundary layer (boundary layer at $x=0$), respectively. 
For this figure as well as figure (\ref{fig:phases2}), 
the vertical dashed line demarcates similar phases 
with or without LBL. The remaining 
two  dashed curves demarcate two similar phases with 
change in the slope of a boundary layer only. }
\label{fig:phases1}
\end{figure}

 In the low-density phase, as $\gamma$ is increased
 keeping $\alpha$ fixed,  $\tau$-profile continues 
to have an outer solution  
identical to that described above until $\gamma$ reaches the upper 
fixed point branch.  In all these cases,
the boundary condition at $x=1$ is satisfied by a $\tanh$ boundary 
layer.

\subsubsection{Shock phase} 
Keeping $\alpha$ fixed, if $\gamma$ is 
increased  beyond the upper fixed point branch, 
the $\tanh$ boundary layer for $\tau$ 
can no longer satisfy the boundary condition at $x=1$ as 
it saturates to $1-\tau_{\rm out}(x=1)$
before satisfying the boundary condition. 
It is at this point that  the boundary layer
 deconfines from the boundary and enters into the bulk. 
This gives rise to a jump discontinuity (shock) in the 
  density profile and the system enters into a 
shock phase. Right on the phase boundary between the low-density
 and the shock phase, the $\tanh$ boundary layer is such that as 
$\tilde x\rightarrow \infty$, 
it saturates exactly at $\gamma$.  
The phase boundary in the $\gamma-\alpha$ plane 
is thus determined from the condition 
$\gamma=1- \tau_{\rm out}(x=1)$, where 
$\tau_{\rm out}(x=1)$ is a function of $\alpha$. In the shock 
phase, as one increases $\gamma$ further, the  boundary layer moves towards
the $x=0$ boundary. In this case,  
as $\tilde x\rightarrow \infty$, the boundary 
layer approaches    
another   outer solution part which extends until  $x=1$ 
and satisfies the boundary condition there.

The emergence of a shock phase in the phase diagram has similar 
origin as that observed earlier for a 
single lane ASEP with Langmuir kinetics 
\cite{parameggiani,evans3,smsmb}. 
However,  due to the interplay between the two densities $\tau$ 
and $\sigma$, we have a much more intriguing situation here.
It needs to be noted that 
while the $\tau$-profile  here can support a shock ( a boundary layer 
located anywhere in the interior of the lane), 
 a boundary layer for  $\sigma$  is constrained to be only near $x=1$. 
As seen in figure (\ref{fig:shock}), a jump discontinuity in 
the $\tau-$profile  leaves its signature 
on  the $\sigma$-profile by producing a discontinuity in  the 
slope of the outer solution of $\sigma$.
In order to understand this, we  study the differential equation 
for  $k(x)=\frac{{\rm d}\sigma(x)}{{\rm d}x}$. Rewriting (\ref{4}) as a 
differential equation for $k(x)$ and then introducing
 the scaled variable
$\tilde x$,   we have 
\begin{eqnarray}
\frac{{\rm d}k(\tilde x)}{{\rm d}\tilde x}-\frac{v}{D_\sigma}k(\tilde x)
 +\left[\frac{\Omega_{{\rm d}}}{D_\sigma}\tau_{\rm in}-
\frac{\Omega_{{\rm a}}}{D_\sigma}\sigma_c(1-\tau_{\rm in})\right]=0. \label{discn_sigma}
\end{eqnarray}
We have replaced $\tau$ by the inner solution $\tau_{\rm in}$ 
since we are interested in probing the region over which the shock 
appears in the $\tau$-profile.
In order to obtain a simpler equation, we have chosen  a constant 
value for $\sigma$ 
 ($\sigma=\sigma_c$) over the narrow region of the shock. 
The complementary solution of the differential equation is 
 \begin{eqnarray}
 k(\tilde x)=c_1 \exp[\frac{v}{D_\sigma}\tilde x].
  \end{eqnarray}
This solution diverges exponentially as
  $\tilde x\rightarrow \infty$ and does not saturate to a finite 
value as required.  This leads us to
 choose $c_1=0$. It, therefore, 
appears that  only the particular integral, 
gives the solution for the slope. 
Since the shock is  a $\tanh$ kind solution of the boundary layer, 
we insert the $\tanh$-solution for $\tau_{\rm in}$ 
in (\ref{discn_sigma}) and solve it numerically by supplying 
values of all necessary parameters such as  $k_1$, $\tau_0$, 
that are present in the $\tanh$ type solution for 
$\tau_{\rm in}$ and the value of 
$x_c$ at which the shock in $\tau$ is formed. The numerical values 
of all these parameters are obtained from   direct  MATLAB solution 
of the full differential equation under boundary 
conditions that give rise  to a shock in the $\tau$-profile. 
 For example, for $\alpha=0.45$,
 $\gamma=0.8$ and $\Omega_{{\rm d}}=\Omega_{{\rm a}}=0.2$, the MATLAB solution
 shows that the shock in 
$\tau$ is formed at $x=x_c=0.272$ with 
$k_1=-0.065$ and $\tau_0=0.612$. 
These numbers are plugged in the $\tanh$ solution 
in (\ref{inn-tau-sol}) 
which is then used to obtain the particular  integral
 for $k(\tilde x)$ numerically using Mathematica.  

The particular integral for the slope obtained using Mathematica
is compared with the slope obtained from the full numerical 
solution of the differential equations using MATLAB (see figure 
(\ref{fig:shock}b)). 
The mismatch between the solutions away from the center of the 
shock is possibly due to the approximation of choosing 
a constant value of $\sigma$. From 
the entire exercise, one may conclude that around the shock, 
the $\sigma$-profile is described  by the particular integral 
of the complete differential equation
(\ref{4}).

\subsubsection{High-density (HD) phase}
  
  The high-density phase emerges as the shock 
 reaches the $x=0$ boundary with further increase in $\gamma$. 
Hence, right on the phase boundary between the shock
 and the high-density phase, the $\tanh$ type boundary layer saturates 
to $\alpha$ at $x=0$. The other end of the 
 $\tanh$ solution 
must saturate to the outer solution which finally extends up to 
$x=1$ boundary and satisfies the boundary condition at $x=1$. 
Thus right at the phase boundary between 
 the high-density and the shock  phase, 
 the outer solution for $\tau$-profile, 
satisfying  the boundary condition at $x=1$, 
has a value $1-\alpha$ at $x=0$. As always 
is the case, the outer solution for  $\sigma$-profile 
satisfies  the boundary condition at $x=0$. Since, in this phase, the 
boundary layers for $\tau$ and 
$\sigma$ are located at two opposite boundaries, the strategy 
to find the phase boundary needs to be altered slightly. 
We obtain the value of $\tau_{\rm out}(x=1)$ 
by solving the 
 outer equations for $\tau$ and $\sigma$ with 
boundary conditions $\tau(x=0)=1-\alpha$ and $\sigma(x=0)=\alpha$. 
The condition $\gamma=\tau_{\rm out}(x=1)$
finally decides the phase boundary
 between the shock and the high-density phase.

\subsection{ $\alpha>0.5$}
\subsubsection{Low-density phase}
Let us first consider the situation where $\gamma<0.5$.
The flow  trajectories on the  fixed point diagram 
 suggest that there exists  only one option through  
which the density profile is able to satisfy the boundary condition 
at $x=0$ and finally attain a value less than $0.5$. 
This option involves having a boundary layer at $x=0$ represented 
by the vertical line (B) in 
figure (\ref{fig:fixedptfig}). The 
boundary layer at $x=0$ thus saturates 
to $\tau=0.5$ and an outer solution 
starting from $\tau=0.5$ at $x=0$ continues 
with a negative slope  until $x=1$. The boundary 
condition at $x=1$ is satisfied by a boundary layer.  
Depending on the value 
of the outer solution at $x=1$ and the value of $\gamma$,
 the boundary 
layer has a negative or a positive slope  
(See figure (\ref{fig:ldlblandrbl}) for a 
representative $\tau$-profile with a boundary layer at $x=1$ 
with positive slope). 
\begin{figure}[htbp]
\centering
   \includegraphics[width=.45\textwidth]{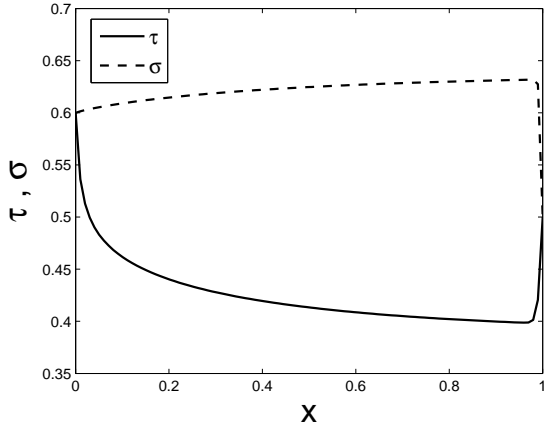}
\caption{ Representative  density profiles in the low-density phase 
for $\alpha=0.6$ and $\gamma=0.5$. 
The boundary layer at $x=1$  has a positive slope.
 Other parameter values are same as those in figure 
(\ref{fig:samplefig}). }
\label{fig:ldlblandrbl}
\end{figure}
Similarly to the 
low-density phase for $\alpha<0.5$, boundary layers 
in this case also will be represented by  vertical 
lines   (A) or (A') of the fixed point diagram.
Consequently,  the phase boundary between  phases
 IVA and IVB is  determined by 
 $\gamma=\tau_{\rm out}(x=1)$.
Since $\tau_{\rm out}(x=1)$ is determined  
simultaneously by solving the outer equations for 
$\tau$ and $\sigma$ with the conditions 
$\sigma(x=0)=\alpha$ and $\tau(x=0)=0.5$, 
$\tau_{\rm out}(x=1)$ is a function 
of $\alpha$.

\subsubsection{Shock Phase}
Keeping $\alpha$ fixed, as one increases $\gamma$ further in  
phase IVB, a shock phase emerges again through the 
deconfinement of the boundary layer. The shape of the density profile 
in this phase typically appears as shown in 
figure (\ref{fig:fixedgamma_shock}). 
\begin{figure}[htbp]
\centering
   \includegraphics[width=.40\textwidth]{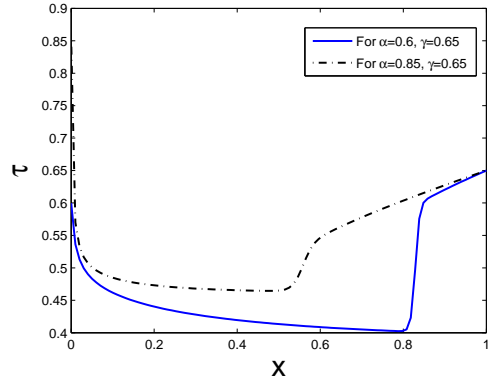}
\caption{  This figure shows how the $\tau$-profile 
changes in the shock phase as the value of $\alpha$ is changed keeping 
$\gamma$ fixed. Parameter values are same as those in 
figure (\ref{fig:samplefig}). }
\label{fig:fixedgamma_shock}
\end{figure}
The mechanism of 
 shock formation is identical to the case of $\alpha<0.5$. 
Here also, the phase boundary between  IVB and the shock 
phase is determined 
from the condition 
$\gamma=(1-\tau_{\rm out}(x=1))$. However,
in this case,  $\tau_{\rm out}(x=1)$ is determined by  
solving simultaneously 
the outer equations for $\tau$ and $\sigma$ 
with boundary conditions 
$\tau(x=0)=0.5$ and $\sigma(x=0)=\alpha$.

For $\alpha>0.5$, the density profiles in the LD 
and shock phases 
have negative-slope boundary layers at $x=0$ saturating 
to $\tau=0.5$. Due to this, the outer solution 
part of the profile always has an effective 
boundary condition $\tau(x=0)=0.5$ even though 
the value of $\alpha$ is changed. The value of
$\tau_{\rm out}(x=1)$, however, 
is dependent on $\alpha$ since, in order 
to find $\tau_{\rm out}$, we have to simultaneously solve
(\ref{out1}) and (\ref{out2}) with the boundary 
conditions $\sigma_{\rm out}(x=0)=\alpha$ 
and $\tau_{\rm out}=0.5$. It is because of this 
dependence that the phase boundaries between phases IVA and IVB 
 and between IVB  and the shock phase acquire a slope rather 
than being horizontal.
In order to give a specific example, let us consider  phase IVA.
In this phase, 
as $\alpha$ is increased keeping $\gamma$ fixed, 
the outer solution of $\sigma$, that satisfies the condition 
$\sigma(x=0)=\alpha$ and continues until $x=1$,  
in general acquires a larger value.  This 
influences the outer  solution for $\tau$ in such 
a way  that $\tau_{\rm out}(x=1)$ becomes 
larger as $\alpha$ increases. 
Since on the phase boundary between IVA and IVB, 
$\gamma$ must be equal to the value 
of $\tau_{\rm out}(x=1)$,
as $\alpha$ increases the value of $\gamma$ also 
 increases leading to a positive slope of the 
phase boundary. This, in conjunction with 
the fixed point diagram, implies that the phase boundary 
between the IVB and the shock phase has a negative 
slope with $\alpha$. The two phase boundaries 
join when $\tau_{\rm out}(x=1)=1-\tau_{\rm out}(x=1)=\gamma$. 
This corresponds to 
$\alpha=\alpha_c=0.987$  for the parameter 
values of figure (\ref{fig:phases1}). 
At 
$(\alpha=\alpha_{c}, \gamma=1/2)$, the height of 
the  right boundary layer vanishes and a  very 
slowly varying inner solution satisfying the boundary condition at 
$x=0$ and saturating  to 
$\tau=0.5$ at $x=1$ is present across the whole lane.  
 As a consequence of the disappearance 
of the right boundary layer, 
the shock phase  terminates 
at $\alpha=\alpha_{c}$ (figure (\ref{fig:phases1})). 
Therefore, right at the edge of the shock phase, the 
density profile has a left boundary layer  
merging to  $\tau=0.5$. 
This is followed by an outer 
solution which continues until
 $x=1$ and satisfies the boundary 
condition at $x=1$ 
(see fig (\ref{fig:shocktermination})). 
Since 
$\gamma > 0.5$ in this figure, 
the outer solution lies in the upper half 
of the fixed point diagram.
\begin{figure}[htbp]
\centering
   \includegraphics[width=.4\textwidth]{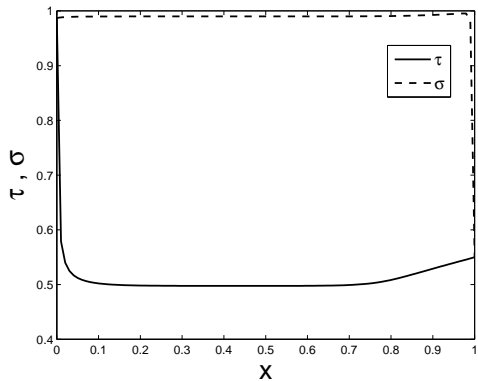}
\caption{ Density profiles for  
$\alpha= 0.987$, $\gamma = 0.55$. Other parameter 
values are same as those in figure (\ref{fig:samplefig}). }
\label{fig:shocktermination}
\end{figure}
Same trend continues if  $\alpha$  is increased 
beyond $\alpha_c$ (shaded region in figure (\ref{fig:phases1})). 
From the fixed point 
diagram and also from the 
discussion below on the high-density phase, 
it becomes apparent that for  $\alpha>\alpha_{\rm c}$, the shock 
phase enters into a high-density phase in which the density 
profile, at $x=0$, has a negative slope boundary layer saturating 
to an outer solution which has $\tau>0.5$ through out.   

For $\Omega_{{\rm d}}=2\Omega_{{\rm a}}$, and  parameter values as shown in figure
 (\ref{fig:phases2}), $\alpha_{\rm c}>1$. 
The termination of the  shock phase is, therefore, not seen in this case.
\begin{figure}[htbp]
\centering
   \includegraphics[width=.45\textwidth]{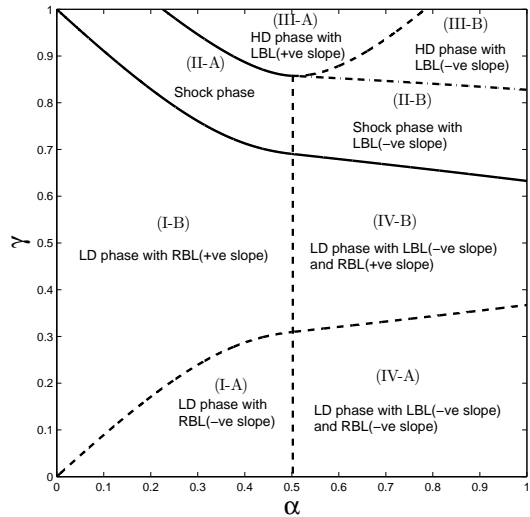}
\caption{Phase diagram for $\epsilon = 0.002$, $D_{\sigma} 
= 0.6$, $v=0.4$, $\Omega_{{\rm d}}=2\Omega_{{\rm a}}=0.2$. As the  
dot-dashed phase boundary is approached from the shock phase, 
the shock height vanishes 
continuously.}
\label{fig:phases2}
\end{figure}

\subsubsection{High-density phase} In the shock phase, 
keeping $\alpha$ fixed at a value sufficiently away from 
$\alpha=0.5$ line, if we increase $\gamma$, we expect 
the shock to be pushed towards the $x=0$ boundary. From the 
continuity point of view, this process should also 
be  accompanied by a reduction of the shock 
height as the density profile in all these cases, already has  a 
negative slope boundary 
layer at $x=0$ merging to $\tau=0.5$ 
(see figure (\ref{fig:fixedalpha_hd})).
 \begin{figure}[htbp]

\centering
   \includegraphics[width=.4\textwidth]{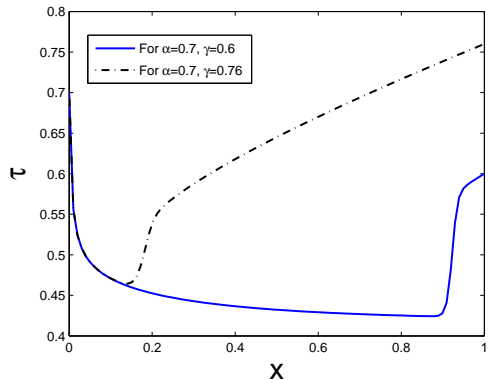}
\caption{ $\tau$-profiles presented here 
show how the  height and the location
of the shock  change as $\gamma$ is changed keeping 
$\alpha$ fixed. Parameter values are same as those in figure 
(\ref{fig:samplefig}). }
\label{fig:fixedalpha_hd}
\end{figure}
 The reduction of the shock height is such that  right 
at the phase boundary between the   
shock and the high-density phase, the shock height disappears. 
 At the phase boundary, the density profile, therefore, 
has a boundary layer of negative slope at $x=0$ followed by an 
outer solution of positive slope 
satisfying the boundary condition at $x=1$. This outer solution as 
before lies completely in the upper half of the fixed point diagram.  
The phase boundary between the shock and the HD phase here is found 
by the condition $\tau_{\rm out}(x=1)=\gamma$ where 
$\tau_{\rm out}(x=1)$ is found by solving simultaneously the outer 
equations for  $\sigma$ and $\tau$ with 
boundary conditions $\tau(x=0)=0.5$ and 
$\sigma(x=0)=\alpha$. 

The phase boundary again has an $\alpha$ dependence. We take the 
example of this phase boundary to compare our result with 
single-lane ASEP with Langmuir Kinetics 
\cite{parameggiani}. In the latter case,
the outer equation describing the density profile on the lane, has a form 
\cite{smsmb}
\begin{eqnarray}
 \frac{{\rm d}\tau_{\rm out}}{{\rm d}x}(2\tau_{\rm out}-1)+
\Omega \left(K(1-\tau_{\rm out})-\tau_{\rm out}\right) = 0, 
\end{eqnarray}
where $K=\Omega_{{\rm a}}/\Omega_{{\rm d}}$ and $\Omega=\Omega_{{\rm d}}$. 
Since, in our case,  the outer solution for $\sigma$ 
satisfies the boundary condition $\sigma(x=0)=\alpha$, 
we use the  approximation that 
 $\sigma\approx \alpha$ in (\ref{out1}). 
With this approximation,  the present model is equivalent 
to a single-lane ASEP with Langmuir kinetics with a varying $K$. 
Although there are significant quantitative differences, 
 two HD-shock  phase boundaries, one obtained from 
the approximate single-lane model and other obtained from 
our original model have qualitative similarities.

\section{Summary}
We consider a driven many particle system consisting of two identical 
lanes of finite length. Particles move with biased 
diffusion on one lane termed as  the diffusion-lane. On the other lane,
 referred as the ASEP-lane, particles hop  in a specific direction 
obeying  mutual exclusion principle. Two lanes mutually 
interact through the  exchange of particles. These particle exchange processes 
give rise to a \textquoteleft transverse 
particle flux\textquoteright\  between 
the  lanes. The densities at the boundaries 
of the two lanes are assumed to be controlled by two reservoirs which 
maintain specific densities at the boundaries. For simplicity, we 
assume that the  boundary densities at left and  
right ends  are same for both lanes.  We denote the  boundary 
densities as 
$\alpha$ and $\gamma$ for left and right ends, 
respectively. As the boundary densities are changed, 
the system exhibits boundary induced 
bulk phase transitions. 
Boundary induced transitions in such a 
two-lane system have  been  studied earlier  using a zero 
 \textquoteleft transverse particle flux\textquoteright\   
  condition \cite{tailleur,evans2,vandana}.
In this paper, we study this system  with  nonzero \textquoteleft 
transverse particle  flux\textquoteright.

By studying the average density profiles on the two lanes,
 we obtain the  phase diagram in the $\alpha-\gamma$ 
space. The phase diagram is based on distinct shapes of density 
profiles  across the ASEP-lane. 
In different phases, the density profile on the ASEP lane  
changes significantly 
due to varying locations of the boundary layers and 
values of the bulk densities. On the other 
hand, the density profile on the diffusion-lane 
 does not show significant variation 
with the  boundary densities. The boundary layer
in the profile is strictly confined to the 
right end of the lane. 
Further, this system provides an opportunity to study 
how the presence of a discontinuity (shock) in the density profile  
of the ASEP-lane affects the density profile  on the diffusion-lane.
This is especially interesting  for this particular 
system  since the profile on the diffusion-lane 
cannot have a jump discontinuity 
in its bulk. We show that  the shock in the ASEP-lane 
produces a jump  discontinuity in the slope of 
the density profile on the diffusion lane.

 \end{document}